\documentstyle[12pt,aasms4]{article}
\def\lsim{\lower.5ex\hbox{$\; \buildrel < \over \sim \;$}}
\def\gsim{\lower.5ex\hbox{$\; \buildrel > \over \sim \;$}} 
\def\lax    {\ifmmode{_<\atop^{\sim}}\else{${_<\atop^{\sim}}$}\fi}
\def\gax    {\ifmmode{_>\atop^{\sim}}\else{${_>\atop^{\sim}}$}\fi}
\def\etal{{et al.\/} }
\def\gtorder{\mathrel{\raise.3ex\hbox{$>$}\mkern-14mu
	     \lower0.6ex\hbox{$\sim$}}}
\def\ltorder{\mathrel{\raise.3ex\hbox{$<$}\mkern-14mu
	     \lower0.6ex\hbox{$\sim$}}}

\begin{document}

\title{LITHIUM PRODUCTION IN COMPANIONS OF ACCRETING X-RAY BINARIES
 BY NEUTRON SPALLATION OF C,N,O ELEMENTS}

\author{ Nidhal Guessoum\altaffilmark{1,2,3}\& Demosthenes Kazanas\altaffilmark{1}} 

\altaffiltext{1}{LHEA, NASA/GSFC Code 661, Greenbelt, MD 20771}
\altaffiltext{2}{Universities Space Research Association}
\altaffiltext{3}{Department of Applied Sciences,  College of Technological
Studies, Shuwaikh, Kuwait, 70654}

%\vskip 0.5 truecm
\font\rom=cmr10
\centerline{\rom submitted to Astrophys. J.}

\baselineskip=15pt

\begin{abstract}

We examine the processes which could lead to the observed enhancement
of Li and possibly other light elements (Be, B) in the companions
of a number of X-ray novae. We conclude that one of the most promising
mechanisms is the spallation of CNO elements on the surface of the
companion induced by the neutron flux produced in the hot accretion 
flow onto the compact object. Direct production of the observed Li and 
its deposition onto the dwarf companion seem less likely, mainly because
of the possibility of its destruction in the production region itself
and difficulties in its deposition associated with the configuration 
of the companion's magnetic field. We discuss other potential 
observables of the above scenario.
\end{abstract}

\keywords{accretion--- black hole physics--- radiation mechanisms: 
nuclear reactions--- stars: neutron--- X-rays}

\section{Introduction} 
 
High Lithium abundances have recently been detected in the secondaries
of at least five soft X-ray transients: V404 Cyg, A0620-00, Cen X-4,
Nova Mus (Martin et al. 1992, 1994, 1996), and GS 2000+25 (Filippenko,
Matheson, \& Barth 1995, and Harlaftis, Horne, \& Filippenko 1996). 
Although the measured abundances were not extremely high, ranging from
about $10^{-10}$ (for A0620-00) to about $2 \times 10^{-9}$ (for Cen X-4),
they are considered highly unusual and surprising simply because in such
late-type stars as the secondaries of these compact objects are, 
Lithium should be greatly depleted by means of nuclear burning within the
stars. This association of enhanced Lithium abundance
with the presence of the compact companion, raises then the possibility
of the {\it production} of Lithium, either in the accretion flow 
surrounding the compact object and its deposition on the surface of 
the secondary, or its production at the latter site by processes 
associated with the accretion flow. 

In addition to this development, one of these sources (Nova Muscae)
exhibited during its outburst a prominent  gamma-ray line feature 
at about 450 keV (Goldwurm et al. 1992), which was interpreted as 
either a red shifted electron-positron annihilation feature, 
or possibly a nuclear deexcitation line from Helium-Helium
collisions in the accretion disk (Martin et al. 1994; Rebolo et al. 
1995; Durouchoux et al. 1996; Guessoum, Ramaty, and Kozlovsky 1997).
In the latter reference, to enhance the line emissivity, it was 
assumed that the accreted matter was Helium-rich, such as would be 
the case if the companion star was highly evolved and had lost its 
Hydrogen envelope (e.g. WN Wolf-Rayet). In that work, it was shown that 
the gamma-ray line flux would be sufficiently high to be consistent with
the 1992 observation of Nova Muscae, provided the luminosity of the 
source was high. In addition, it was pointed out that this scenario 
could also explain the high Lithium enrichment observed in such sources,
provided that a sufficient fraction of the Lithium produced was able to 
escape the production region in the accretion disk and be deposited onto 
the companion's surface. Furthermore, this scenario, postulating 
the occurence of nuclear reactions in the hot accretion plasma, led to 
additional predictions: a) a delayed gamma-ray line at 478 keV should 
be observed upon the decay of $^7$Be into $^7$Li$^*$, knowing that 
$^7$Be has a lifetime of 56 days, and b) that an unusual 
$^6$Li/$^7$Li should be measured in these cases, the ratio 
varying highly (between about 10 and 1) depending on the temperature
of the disk.

A more recent paper (Yi \& Narayan 1997) has independently proposed and
investigated the same scenario, namely the production of Lithium 
in $\alpha-\alpha$ collisions, in an advection dominated accretion 
flow (ADAF; Narayan \& Yi 1994, 1995; Abramowicz et al. 1995; Chen et 
al. 1995), rather than a Shakura-Sunyaev disk. It was argued there
that some of the Lithium nuclei would escape and be deposited onto the
surface of the secondary, enriching the observed abundance. 
This model has the advantage of keeping the ion temperature much higher 
than the electron temperature, thereby allowing the necessary nuclear 
reactions to proceed more efficiently. 
The model is also convenient from the technical point of view,
as it provides simple analytic expressions for the 
radial dependence of the temperature and the
density of matter in the accretion flow as a function of 
the viscosity parameter $\alpha$ and the macroscopic parameters of the 
flow (mass of the compact object and accretion rate).

While important in its basic setting and concepts, the treatment of 
Yi \& Narayan (1997) did not take into account several processes which
can drastically modify the resulting Li production:  
First and foremost, the authors did not take into account the {\it
breakup} of both Lithium and Helium by protons (Guessoum and Gould 1989) in
the hot accretion flow. This point is of the utmost importance since both 
these processes will compete strongly with the Helium-Helium reaction 
which produces $^7$Li. In this respect, one should note that the 
breakup of $^7$Li (by protons) into 2 $\alpha$'s is an {\it exothermic}
reaction, having no threshold energy to overcome, while at the same time,
the breakup of Helium has a higher cross section than the 
$\alpha-\alpha$ reaction which produces the Lithium although its 
threshold is slightly ($\lsim 6 \%$) higher (see table 1);
 hence the same high nucleon temperatures which allow the 
production reaction to take place are more likely to destroy both 
the products and its reactant constituents. Moreover, the capture of 
a proton by $^7$Li (giving $^7$Be and a neutron), has a very low 
threshold and a very high rate. This latter reaction is normally 
in equilibrium with its inverse reaction (the capture of a neutron 
by a beryllium nucleus giving $^7$Li and Hydrogen), but in the 
present situation, the protons in the accreting matter are presumably 
much more abundant than the neutrons, and so $^7$Li is much more
readily destroyed. The same is true also for $^6$Li which is rapidly 
broken up by the protons.
Secondly, the authors did not consider the use of a Helium enriched 
accreting matter which could have alleviated some of the problems
outlined above: indeed, our calculations show that had one used a 
50\% H - 50\% He (by mass) composition instead of their 75\% H - 
25\% He, it would have resulted in an enhancement in Lithium production
by a factor of 4 or more depending on the parameters. Finally, 
the analytic expressions used for the cross sections of the alpha-alpha 
production of Lithium were too approximate. 
The above points will be discussed in more quantitative manner in section 3.

The deficiencies in the Yi \& Narayan (1997) treatment mentioned above, are
not the only impediments to materializing the proposed scenario for the
enhancement of the Lithium abundance on the surface of the companion.
There is in addition the issue of escape and deposition of the Li nuclei 
onto the companion's surface.  It is difficult indeed to see how these
nuclei, charged and relatively heavy, will overcome both the gravitational
barrier and the magnetic fields surrounding the compact object. 
In thermal equilibrium, heavier nuclei (atoms) tend to be 
preferentially favored for capture rather than escape. One could  
invoke the outflows, known as jets, to explain the escape of the 
nuclei from the accretion flow (it should be noted that the mechanism 
of jet production is not at present well understood); in such a case 
one would have to argue, additionally, that the jet beam will be 
intercepted by the companion star, a rather unlikely proposition given the
general belief that the jet direction is perpendicular to the plane of 
the binary orbit. Last but not least, should the outflowing nuclei be
intercepted, one way or another, by the companion star,
it is by no means certain that they will be deposited on its surface;
this depends strongly on the geometry of the magnetic field lines
of the companion star. It is well known from solar flare observations that
deposition of particles onto the solar surface proceeds only through
loss cone deposition, which allows only a small fraction of the available
particles to hit the solar surface, while galactic cosmic rays, for example,
never make it to the solar surface, otherwise they would produce an
observable flux of $\pi^{\circ}$-decay gamma rays. 

For all the above reasons, and because we consider the correlation of 
enhanced Li abundance sites with the stellar companions of X-ray novae as  
significant, indicative of the presence of a hot ($T_i \sim 30$ MeV) 
accretion flow in which nuclear processes are important, we examine closer 
a variance of the scenario proposed by Yi \& Narayan (1997). Our scenario
retains the general features of ADAF; however, rather than insisting on 
producing the required  Li in the accretion flow, we consider instead the
production of neutrons in the same setting, by the dissociation of Helium
or heavier nuclei. The production of Lithium, then, takes place 
on the surface of the companion, by spallation of CNO nuclei there, 
induced by the neutrons flying from the hot accretion plasma onto the 
compact object.

Neutrons are relatively easy to produce in ADAFs (either in $p-\alpha$
or $\alpha-\alpha$ reactions), and a substantial fraction of them 
are usually produced with 
sufficient energy to allow their escape from the gravitational 
potential of the production site. Furthermore, being neutral, their 
escape, and also their deposition on the surface of the 
companion, is not affected by the presence of magnetic
fields, and it is expected to be roughly isotropic. These properties
allow, then, one to {\it calculate} rather reliably
their escaping fraction and flux onto the surface of the companion, where
the Lithium can be produced. Therefore, the assumption that the 
observed Li is synthesized on the surface of the companion, by neutrons 
produced in the hot accretion flow onto the compact object, greatly 
reduces the uncertainty of the problem, to effectively the viscosity 
parameter $\alpha$ of the ADAF (for a given accretion rate and mass 
of the compact object). It is worth noting that our basic assumption
can be potentially confirmed through direct observation: If the 
observed Li enrichment is indeed due to CNONe spallation by 
neutrons emitted by the compact
object, it should be accompanied by a similar enhancement in the Boron 
and Beryllium abundances which are also by-products of the same reactions.
We shall come back to this point later.

Though difficult to ascertain their presence, free neutrons are generally
present in astrophysical settings in which processes involving nuclei of 
energies $\gsim 10$ MeV are at work. However, their presence is of little
observational consequence if they do not undergo any substantial interaction 
before they decay and/or if their energy density is small compared to that
of the associated plasmas. On the other hand, if they do interact before decay,
and/or their energy densities are not negligible, then their effects can be 
significant. Thus, the neutron 2.223 MeV proton-capture line has been observed 
during solar flares, while neutrons have also been observed directly by 
terrestrial neutron detectors, following such events (for a review see 
Mandzhavidze \& Ramaty 1993). Furthermore, because
neutron propagation is unaffected by the presence of magnetic fields 
and because, by virtue of their production mechanism, they can have 
energies and energy densities comparable to those of the producing plasmas, 
they are excellent agents of energy, momentum and angular momentum transport
in astrophysical plasmas. Their transport properties and effects have been 
considered a number of times in the past literature. 
Thus, Kazanas \& Ellison (1986b) appealed to the production of 
relativistic neutrons from the photodissociation of very energetic 
($10^{15}$ eV) He nuclei, as means of transporting accretion energy from 
the bottom of an accretion flow near a neutron star to its eclipsing 
companion, in order to produce $\gamma-$rays in phase with the eclipses'
ingress and egress. Giovanoni \& Kazanas (1990) and Contopoulos \& Kazanas 
(1995) considered relativistic neutrons as a means of powering the relativistic
jets observed in Active Galactic Nuclei (AGN). Atoyan (1992) has considered 
the effects of neutrons produced near the compact object in AGN on the 
accretion flows which power them, while Guessoum \& Kazanas
(1990) have explored their potential importance as the viscous agents of  
accretion flows onto compact objects.  

The present treatment is much more limited in its scope: It assumes 
the presence of  an ADAF similar to that of Yi \& Narayan (1997) in order to 
model the flow onto the compact member of the X-ray binary (nova),
and examines in detail its consequences within the
scenario involving the production of neutrons outlined above. 
In \S 2 we briefly describe the background flow, its scalings and range
of the parameters involved. We also discuss and give a list of the 
most prominent reactions associated with the interaction of nuclei in the
flow and in particular the production of neutrons. In \S 3 we discuss the
details of the neutron production, as well as of several other nucleids, 
in the ADAF along with the expected gamma-ray line fluxes.
In \S 4 we discuss the production of Li on the surface of the companion 
given our computed flux of neutrons and in \S 5 the results are summarized and 
discussed.

\section{ The Accretion System}
\subsection{Accretion Model and Parameters}
%\medskip

The nuclear reactions relevant to our problem and to the production of gamma
rays necessitate energies of about 20 MeV per nucleus; therefore, 
these processes can potentially take place only in the inner regions 
of an accretion flow,  provided additionally, that the ion temperatures
could reach values of at least several MeV. The latter condition, therefore,
requires that the emitting plasma be optically thin, since the standard
optically thick -- geometrically thin Shakura--Sunyaev (1973) disk is 
much too cool for the required nuclear reactions to take place. The Advection
Dominated Accretion Flows (ADAF) of Narayan \& Yi (1994), which allow for
ions with roughly virial temperatures, provide a natural setting for 
our calculations. 
We therefore consider the regions around the compact object between
about $100 \ R_S$ (where $R_S$ is the Schwarzschild radius of the object, 
$R_S = 2GM/c^2$) and either the surface of the neutron star or the horizon 
of the black hole, depending on the nature of the compact object. In order for our 
treatment to remain generally applicable to both possibilities, we  
assume a mass of $1 \ M_{\odot}$ for the compact object, unless a specific
value is considered.  Because in the present treatment we are only interested in 
the nuclear reaction rates and not in the entire emitted spectrum, the value
of the electron temperature does not affect our results, which thus depend 
solely on the value of the ion temperature T$_i$. 

In the present investigation, and for the reasons discussed above, we use the ADAF 
prescription to determine the dynamics and thermal properties of the 
accreting fluid; hence the ion temperature T$_i$ 
is given by the expression (Yi \& Narayan 1997)
$$ T_i = 1.11 \times 10^{12} \, r^{-1} {\rm K} = 95.68 \, r^{-1} \; {\rm MeV}
\; \simeq 0.1 \, r^{-1} \, m_p c^2
\eqno(1)$$

\noindent where $r$ is the dimensionless radial variable $r = R / R_S$.

Within ADAF, the density in the hot plasma is given by an analytic
expression derived from mass flux conservation; this has a form very
similar to that for spherical accretion and it is given by (Yi \& Narayan
1997) 
$$ n = 6.4 \times 10^{18} \; \alpha^{-1} \; m^{-1} \; \dot m \; r^{-3/2} \;
{\rm cm}^{-3}\eqno(2)$$

\noindent where $m$ is the dimensionless mass measured in units of a 
solar mass ($m = M / M_{\odot}$), $\dot m$ is the dimensionless accretion 
rate measured in units of the Eddington accretion rate $\dot m = \dot M
/ \dot M_{Edd} = \dot M \, c^2 \, /L_{Edd} = 
{\dot M} / {1.4 \times 10^{17} m}~~ {\rm g/s}$, and where,
following Yi and Narayan, we have taken $\beta$ (the ratio of gas pressure to
total pressure) to equal 1/2.

The largest, by far, part of the kinetic energy associated with 
the inflow is carried 
by the hot protons which provide most of the accretion inertia. Their 
energy is given off to the electrons via Coulomb collisions (or some other 
type of plasma interaction) and then transferred efficiently to photons to 
produce the observed radiation. In this scenario, therefore, the 
observed luminosity depends on the rate of energy transfer from 
the hot protons onto the electrons (provided that the associated time scale
is longer than the local free-fall time). If $\sigma \ v$ is the rate of the
relevant reactions, the total luminosity of the system is given by $L 
\simeq \int n(r)^2  \sigma \;v \; kT_i \; dV$ where 
$dV = 4 \pi R^2 dR = 4 \pi \; R_S^3\;  r^2 dr \simeq 3.4 \times 10^{17}\;  
m^3 \; r^2\;  dr~~~{\rm cm}^{3}$ is the volume element in the accretion flow. 
The resulting luminosity, then, will be

\begin{eqnarray}
\setcounter{equation}{3}
L & \simeq & 4.4 \times 10^{38} {k T_i \over m_pc^2} 
{\sigma \;v\over \sigma_T \;c}
\; \alpha^{-2} \; m \; {\dot m}^2  \nonumber\\
  & \simeq & 4.4 \times 10^{37} {1 \over r_{in}} 
{\sigma \;v\over \sigma_T \;c}\; \alpha^{-2} \; m \; {\dot m}^2
~~~~ {\rm erg \; sec}^{-1} 
\end{eqnarray}
\noindent where in the last step the value of $k T_i$ was substituted
using equation (1), with $r_{in}$ denoting the innermost radius for which the 
scaling of equation (1) holds and $\sigma_T$, the Thomson cross section, 
was introduced to 
provide a normalization of the reaction rate $\sigma \,v$ to $\sigma_T \,c$.
Assuming $m$ to be in the range of 1 - 10, the entire uncertainty of the 
problem is in fact reduced to the value of $\alpha$ (following Narayan
\& Yi 1994 we consider $\alpha = 0.3, \; 0.1$) and to a choice for the
accretion rate $\dot m$ or equivalently $\dot M$, which can be obtained 
from the observed total luminosity of the system; in the present note we
will consider that to be in the range of $10^{-12} \ M_{\odot}/yr$ and 
$10^{-8} \ M_{\odot}/yr$.

Finally we note that the system is taken to be in steady-state,
whereby the accretion and in-fall are constant in time, and thus the
neutron production and gamma-ray emission are steady.

%
%\bigskip
\subsection{The Nuclear Reactions}
%\medskip
%
Assuming an initial composition for the accreting plasma, usually some 
proportion 
of Hydrogen and Helium, we set up a network of nuclear reactions which
the ions (the primary protons and alphas as well as the secondary species)
can undergo, including those leading to de-excitation gamma-ray photon
production; we then let the plasma evolve over the dynamical time scale $t_d$ 
with its temperature following values given by Equation (1) as it
sinks in the gravitational well.  The main reactions in our simulations are 
displayed in Table~1 along with the center-of-mass threshold energies 
necessary for these reactions to proceed.  These energies serve as a rough
indicator of the relative importance of the various channels (since
the cross sections are approximately of the same magnitude).

The variations in the abundances of the nuclear species are treated through
a reaction rate formulation, using an expression for binary processes that 
is correct for non-relativistic energies:
$$ r_{ij} \equiv {dN_{ij} \over {dV\ dt}} = { 1 \over {1 + \delta_{ij}}}
\int\int \sigma(v_{ij})\ v_{ij} \ dn_i \ dn_j \, , \eqno(4) $$
where $i$ and $j$ are the interacting nuclei in a particular reaction,
$\sigma (v_{ij})$ its total cross  section, 
$v_{ij}$ is the relative velocity of the two species, $\delta_{ij}$ is the 
Kronecker delta and $dn_i \ , \ dn_j$ are the
distribution functions of the particles in velocity space.  We assume
that the timescale for thermalization among the ions is much shorter
than that for the nuclear reactions or the Coulomb collisions, so
that the particles' distribution functions are always Maxwellian (see
Guessoum and Gould 1989 for a discussion of this point).  We use
experimentally-measured cross sections to obtain the reaction rate of
each nuclear reaction as a function of the ion temperature (references 
for the cross sections are given in Guessoum and Gould 1989; Guessoum and
Kazanas 1990; see also Glagola et al. 1982). We then use
a simple numerical scheme of explicit finite-differencing to follow
the abundances of the various species as well as the photons produced, 
as the ions sink in the gravitational well of the compact object with
the temperature changing according to equation (1), 
until they are removed from the plasma at $t = t_{d}$.  This
calculation is performed for various values of the model parameters:
$\dot M = 10^{-12}, \; 10^{-11}, \; 10^{-10}, \; 3 \times 10^{-10}, \; 
10^{-9}, \; 3 \times 10^{-9} \;$ and $10^{-8} \; M_{\odot}/
{\rm yr}$; $\alpha = 0.3 $ and $ 0.1$; $M = 1, $ and $ 10$; a Helium 
fraction  $Y = 10\%, \; 30\%, \; 90\%$, etc.

Concerning the gamma-ray lines, since we assume the plasma not to contain
any nuclei heavier than Helium, then the only lines of interest in our
case are the following:
\begin{itemize}
\item 0.429 MeV resulting from the de-excitation of
${}^7$Be$^*$, which is produced mostly in $\alpha-\alpha$ collisions;
\item 0.478 MeV resulting from the de-excitation
of ${}^7$Li$^*$, which is produced mostly in $\alpha-\alpha$ collisions,
plus a contribution from $^7$Be, which, when it decays, produces
${}^7$Li$^*$ 10\% of the time;
\item 3.561 MeV resulting from the de-excitation
of ${}^6$Li$^*$, which is produced mostly in $\alpha-\alpha$ collisions;
\item 2.224 MeV resulting from the capture of
neutrons on protons, where the neutrons and some of the protons are produced 
in the breakup of alphas.
\end{itemize}
%
%\vskip 1 truecm
\section{Production of Neutrons, Nuclear Species, and Gamma-Rays}
%\bigskip
%
\medskip
\subsection{Neutrons}
\medskip
Once the reaction rates for the network of nuclear reactions have all been
calculated, the numerical scheme is run for each set of parameters, 
assuming that the temperature profile is constant in time and given by 
equation (1). The normalization of the neutron and nuclear line photon 
production rates are then given by integrating the reaction rates (computed
as discussed above) multiplied by the square of the local plasma
density,  over the volume of the flow within which the plasma temperatures
are above the reaction energy thresholds, indicated in the previous section. 
This prescription leads to the following value for the total rate of 
production of a given species (be that an ion or a photon)

$$R_{tot} \simeq 3 \times 10^{41} \; {\sigma \;v\over \sigma_T \;c}
\; \alpha^{-2} \; m \; {\dot m}^2\; lnr_M ~~~~ {\rm sec}^{-1} \eqno(5)$$

\noindent where $r_M$ is the maximum radius within which the relevant 
reaction is energetically allowed.

One should note that the radial density dependence of ADAF guarantees 
a rather weak explicit dependence of the rates on the outer edge of 
the reaction volume; most of the dependence is implicit through the 
reaction rate $\sigma \,v$. It
also indicates that the total reaction rate is proportional to the mass of 
the compact object and to the {\it square} of the normalized accretion 
rate $\dot m$ (rather than being proportional to it; this is a property
of ADAF and generally any accretion flow in which the emitted luminosity 
is the result of binary interactions within the flow; see e.g. Kazanas \& 
Ellison (1986a)), or in view of equation (3), proportional to the bolometric 
luminosity $L$.

Tables 2a and 2b present the results of these calculations, showing the 
rates of neutron production and escape for various values of the viscosity
parameter $\alpha$ and the mass accretion rate $\dot M$. All the results
shown correspond to $ M = 1 M_{\odot}$, and $R \approx 100 \ R_S 
\approx 3 \times 10^7$ cm. 
Table 2a corresponds to a solar initial composition (i.e. 90\% H and 10\% 
He by number) of the plasma, while Table 2b corresponds to a 90\% He, 10\% H 
composition, appropriate for a He-rich companion.

It is interesting to note that the resulting neutron fluxes depend quite 
strongly on the 3 parameters considered, namely $\alpha$, $\dot M$, and 
the initial composition of the plasma.  For the smallest values of $\dot M$,
the neutron flux scales roughly as its square (and also as $\alpha^{-2}$), 
as demanded by the  corresponding scaling of the emission measure. However,
for higher values of this parameter the flux saturates as all the available
Helium nuclei break up and release their neutrons. These saturation values 
are in fact quite high when one considers that they correspond to the fluxes
of the {\it escaping} neutrons.

%
%\bigskip
\subsection{Li, Be, ${}^3$He}
%\medskip
%
The numerical scheme described above can also keep track of the variation 
of the abundances of all the nuclear species of relevance, such as ${}^7$Li,
${}^7$Be, ${}^6$Li, ${}^3$He, etc. One can compute the final abundances of 
these isotopes for each set of parameters and compare them to each other or to
neutron abundances in order to assess the efficiency of the production of each 
nucleus compared to that of neutrons.

Our results are presented in Tables 3a and 3b where we show the relative abundances 
of neutrons and ${}^7$Li, for various values of the viscosity parameter $\alpha$ 
and the mass accretion rate $\dot M$. As for Tables 2a and 2b the results shown 
here all correspond to $ M = 1 M_{\odot}$ and $R \approx 100 \ R_S \approx 3 
\times 10^7$ cm. Again Table 3a is for a solar initial
composition (i.e. 90\% H and 10\% He by number) of the plasma; Table 3b is
for a 90\% He and 10\% H composition.

It is interesting to note that final neutron abundances are much higher than
those of Lithium, although the values given in these tables do not take into
account the eventual escape of the neutrons, and therefore these numbers are 
useful only for comparative purposes.

We have also estimated the error made by using the approximate 
cross sections of Yi \& Narayan (1997) for the 
production of Li by He-He collisions, instead of the 
experimentally measured cross sections we have used (Glagola et al. 1982). 
Their expression for the cross section overestimates 
the actual data by about a factor of 2 at low energies ($8.5 MeV 
< E \lsim 40 MeV$) and by a factor of about 10 at higher energies 
($ E \gsim 50 MeV$). This then translates into an overestimate of 
the Li abundance achieved in the disk by as much as an order of magnitude.

Moreover, we have calculated the effect of Lithium and Helium breakup in the
flow on the final Li abundance by running our numerical code cases with or 
without the destruction reactions. 
We find that for low accretion rates ($\dot m \lsim 10^{-3}$) the difference
is small, but for substantial rates ($\dot m \gsim 10^{-3}$) neglecting the
breakup results in an overestimation of the abundance of $^7$Li by a factor
of up to 30 at the highest end of the accretion rate range. This sensitivity 
on the accretion rate is to be expected: Li production, and generally all 
binary reactions in an ADAF, proceed vigorously when the inverse reaction
rate becomes comparable or shorter to that of free fall,
i.e. for large $\dot m$. However, while the Li production rate depends
on the square of the He density (and thus is proportional to $\dot m^2$),
its destruction rate is proportional to the product of the Li and H densities,
i.e. to  $\dot m^3$, leading to the computed much reduced final Li abundance
at high $\dot m$. 

%
%\bigskip
\subsection{Gamma-Ray Fluxes}
%\medskip
%
As mentioned in the introduction, one can also calculate, through the same
numerical program, the fluxes of the nuclear gamma-ray lines emitted in
the accretion disk. For this, one needs only the cross section for the
excitation of $^7$Li,$^7$Be, and $^6$Li either in $\alpha-\alpha$ collisions
or in collisions of $^7$Li,$^7$Be, and $^6$Li with protons;  for lack of 
experimentally measured cross sections, we disregard the latter possibility.

Table 4 shows the sum of fluxes of the 0.478 keV and 0.429 keV lines, since 
they combine to form a single broad feature (their Doppler widths being of the 
order of $\sim$ 100 -- 300 keV). The fluxes are again shown for various values 
of the viscosity parameter $\alpha$ and the mass accretion rate $\dot M$. 
However, because the fluxes are extremely low for normal abundances, we only 
show here the results corresponding to a 90\% He and 10\% H composition.
The other parameters retain the same values.

We note that the fluxes obtained here are much lower than those in Guessoum,
Ramaty, \& Kozlovsky (1997), first because the accretion rates assumed here
are very low ($10^{-12}, \; 10^{-11}, \; 10^{-10}$, etc.), and second because
the density of matter in the ADAF model used in the present investigation is
much lower than that used in the above reference. 

\vskip 1 truecm
\section{Production of Li and Other Elements on the Companion Star}
\bigskip
The neutrons having been produced in the accretion disk, 
they can escape rather easily. The escape is handled the same way 
it was done in Aharonian \& Sunyaev (1984) and Guessoum \& Kazanas (1990), 
that is, one simply needs to evaluate that fraction by counting the
neutrons that have enough energy to overcome the gravitational potential
energy, while taking into account the possibility of elastic neutron-nucleon
collisions, which will impede the neutron escape. For simplicity we assume the
neutrons to have a thermal distribution of temperature equal to that of the protons and the $\alpha$'s, the
temperature being that of the shell were they are produced (i.e. between 10
and 100 MeV); we believe this assumption to be reasonable, especially at high accretion rates and/or temperatures less than say 50 MeV, when the time 
scale for thermalization becomes small compared to the free fall time 
scale. Assuming further that the neutrons are produced and fly out
isotropically, one also needs to average over angles. Escape fractions are
found to range between 10 - 30 \%.

The neutrons could in principle also decay in flight, but the sources we are
interested in have companion stars sufficiently close to their accreting 
compact objects (the orbital periods are about a few hours, indicating 
separations of order of $10^{11}$ cm) that the beta-decay of neutrons is relatively unimportant.

The above considerations indicate that one can calculate in a straight
forward way the flux of neutrons impinging on the surface of the 
star companion to the compact object. The neutrons will then undergo
two types of processes: inelastic collisions which degrade their energies 
leading to their eventual capture by hydrogen, and spallation of the heavier
elements (C, N, O, Ne, etc.). To compute the amount of Li, Be, B produced by
the latter reactions one needs to know the relative abundances of the heavier 
elements in the atmosphere of the companion star, which can in fact vary 
significantly depending on its evolution stage, its history, etc. 
and though we have raised the possibility of a He-rich star atmosphere,
in the following calculations we will limit ourselves to the solar 
composition case.

The nuclear spallation reactions which the neutrons can induce are 
presented in Table 5, along with the threshold energies for each channel.
As previously noted, these spallation reactions will not only produce
Lithium, but also Beryllium and Boron. It is also possible in principle to
break the latter species into Lithium and Helium, but this will take a 
much longer time and will be presently ignored.

Finally, in order to compute the resulting light element abundances one needs 
the cross sections for these specific reactions and channels. Unfortunately, 
experimentally-measured values for these are not available in the literature,
so one is forced to use estimates. In order to remain on the conservative side, 
we assume the neutron-induced spallation reactions to have half the 
proton-induced spallation reactions cross sections;  the latter have been 
compiled by Ramaty, Kozlovsky, Lingenfelter, and Reeves (1997). These cross 
sections typically rise very rapidly with energy near threshold, then take 
constant values usually between 10 and 30 mb.

We can now compute the fraction of CNONe nuclei which will be converted 
into Li when exposed to a given  
neutron flux. Assuming a value $R_n$ for the total neutron production
rate, $D$ for the separation between the compact object and the companion,
$X_{C,N,O,Ne}$ for
the abundance of the corresponding C, N, O or Ne on its surface, and 
$\sigma_{sp}$ the spallation reaction cross section, the fraction $f$ of
these elements which will be spalled after exposure to a neutron flux
corresponding to these values of the relevant parameters for a time 
$\tau$ will be
$$ f = {R_n \over 4\pi D^2} \sigma_{sp}\tau X_{C,N,O,Ne}~~~~. \eqno(6)$$
Assuming as  fiducial values for the neutron production 
rate the value $R_{39} = (R_n/10^{39})$ neutrons/second, for the
time scale $\tau_0 = (\tau/ 1$ year), and for the  distance $(D_{11} = D/ 
10^{11}$ cm) we obtain 
\footnote {The semi major axis of most of the systems discussed 
in the introduction is a few times this value, while it is an order of 
magnitude larger for V404 Cyg} 

$$f  \simeq 10^{-5}  {R_{39} \over {D_{11}}^2} \tau_0 X_{-3} \; . 
\eqno(7)$$

The above estimate indicates that in order to enhance Li to the observed
values one needs to spall roughly $10^{-7}-10^{-6}$ of the available
C, N, O, Ne nuclei (the value of $X$ in the above equation has been 
normalized to the sum of the C, N, O and  Ne abundances). 
It appears from the above figures that a short exposure 
to the above flux over a short flare lasting roughly a month 
 would suffice to produce the observed
enhancement. Of course, the neutron fluxes are expected to be as high 
as the fiducial value only during times of outburst, which last for 
several months and occur roughly once a decade. 

Yi \& Narayan (1997) have discussed in some detail the question of depletion
of Lithium from the atmosphere of the companion star once it has been 
deposited/produced there. The time scale they have used for this is $\sim
10^7$ yrs (they quote estimates from the literature ranging between $10^6$ and
$10^8$ yrs). Clearly this time scale is much too long to alter the 
conclusions of our model given above. Of course, the more relevant time 
scale might simply be the convection time scale of the companion star's
convective envelope, since that is the time scale at which material is mixed. 
That mixing time scale is of order $\tau_r \sim R_*/v \sim 10^6 R_{11}/v_5$ s, 
where the velocity of convection 
has been taken to be $1$ km/sec. That means however that the entire convective 
envelope will be mixed on similar time scales ($10^6$ s), and thus irradiation 
over the duration of an X-ray outburst (several months) will result in an
enhancement in the Li abundance at about the observed levels throughout
the entire convective envelope. Depletion (destruction) of the Lithium happens 
only through convective overshoot, which allows matter from the envelope 
to penetrate to deeper, hotter regions where the Lithium can be destroyed (Sweigart, private communication), to which the time scale quoted by 
Yi \& Narayan (1997) refers. So, if these outbursts last only a few months 
(and not millions of years), convective mixing will only homogenize a given 
Li overabundance over the volume of the envelope, while convective overshoot 
will not have time to get this Li destroyed.

In our view, a more stringent constraint on the Li abundance built up
comes from the fact that the surface layers of the companion are
gradually removed  to be accreted onto the compact object. 
It is hence useful to estimate the time scale it takes for the exposed 
material on the companion's surface to be removed and accreted onto the 
compact object. If $n(h)$ is the matter density as a
function of depth from the stellar surface, the neutrons will penetrate
to a depth at which $\sigma \int n(h)\; dh \simeq 1$, with $\sigma \simeq
10 ~~{\rm mb} = 10^{-25} ~{\rm cm}^{-2}$. Assuming a radius $R = 10^{11}
R_{11}$ cm for the companion, the mass of the surface layer affected by
the neutrons, within which one expects an enhanced light element
abundance, is 

$$M(h) \simeq 4 \pi R^2 \int n(h) \; dh \simeq 10^{-9} \, R_{11}^2 \; M_{\odot}
\; .  \eqno(8)$$

This amount of mass would require about a year to be removed, 
if accreted at the 
Eddington rate onto a compact object of $m = 1$. It thus becomes apparent
that, within the present model,  there could very well be a variability
in the observed abundances, with the highest abundances  most likely 
to be observed shortly after a major outburst, before the processed 
matter is accreted back onto the compact object.  One might argue that
variation in the observed abundance due to the removal of these surface
layers would not take place because the removed Li would be regenerated
by the neutrons produced in the hot ADAF. However, this is not so, because
while the removal rate is proportional to $\dot m$ the neutron production 
in an ADAF is proportional to $\dot m^2$. Thus, following the vestiges of the present model, one concludes
that monitoring the potential variability of the Li, Be, B abundances 
in X-ray novae systems could provide a long term probe of the accretion 
mode onto the compact objects.

%
%\vskip 1 truecm
\section{Summary and Conclusions}
%\bigskip
%
In the present note we have reexamined in greater detail the scenario
put forward by Yi \& Narayan (1997) for the production of the Li 
overabundance, observed in the companions of the X-ray novae,  
in the associated hot
accretion flow around the compact object. While we think that their 
general notion of a hot ADAF as the fundamental cause of the observed
Li enhancement is basically correct and in
agreement with the apparent association of these objects with the X-ray 
novae companions, we have opted for a  variation of the 
above scenario. In particular, we have proposed that the enhancement of 
the Li abundance takes place on the surface of the late type, low
mass companions of these systems by the high flux of neutrons produced
in the hot ADAF. We believe this to be a more promising scenario as it 
does not suffer from the problems associated with the destruction of 
the Li in the ADAF itself, or the uncertainties of its transport from 
the accretion flow onto the surface of the companion star.

We have also suggested and investigated the possibility that Li abundance 
enrichment would, in such scenarios, be enhanced if the accreted matter were
Helium-rich, such as would be the case if the secondary were a highly evolved
star (like a Wolf-Rayet) with an atmosphere consistently in large part of 
Helium and/or Carbon and Oxygen. This idea is indeed a speculative one, and
there is no direct observational evidence that any one of the secondaries of 
the soft X-ray transients is or resembles a WR star; however, optical studies 
of Cen X-4 have led Chevalier et al. (1989) to suggest that the secondary in 
this case is probably a stripped giant, i.e. one with a rich-He atmosphere. 
It is interesting to note that of the 5 SXT's observed to have high-Li abundances, 
Cen X-4 has the highest such 
enrichment. We should point out here that our calculations 
of the production of Li at the companion star (section 4) assumes a normal, solar
composition of the star ($X_{C,N,O,Ne} \approx 10^{-3}$). This possibility of 
Helium enrichment in the accreted matter could reveal itself in future observations.

Perhaps one of the most interesting consequences of our scenario is 
the prediction of the presence of other light elements, namely 
Be and B, in the same sites where the Li is found (see Table 5). 
Unfortunately,
the most prominent lines of these elements are in the UV
and therefore not accessible with ground telescopes. One of us, (DK)
along with Chris Shrader, have obtained spectra of these objects in 
the UV using GHRS on HST, in search of the Be and B lines, considering 
the Li line to be emitted from the accretion disk itself. However, 
at the time of the observation the UV disk emission from these objects was
negligible and in the meantime it became apparent that the Li line 
emission is actually associated with the companion to the compact object. 
Since the companions to these objects are usually low luminosity, late 
type stars, the UV flux in our observations was too small to
provide the  high S/N measurements required for
such detections. A more favorable prospect for the detection of the Be 
and B lines is to search for them in High Mass X-ray Binaries, 
like Cyg X-1 which are bright in the UV and which are thought to 
contain ADAFs favoring the production of neutrons, similar to those 
discussed above (Yi \& Narayan 1997).  

An additional conclusion of the
present model which has potential observational consequences is the 
possibility of time variability of the observed light element (Li, Be, B)
abundances. One expects these abundances to be highest shortly after 
a major outburst, thus making these epochs the most favorable for their
detection. We hope that this paper will 
stimulate further searches in the UV for the transitions associated 
with Be and B, also for less prominent transitions in the optical
domain, but with the larger telescopes presently available, as well as 
the potential time variability of the Li abundance. 

The prominent neutron fluxes indicated by our model raise, in addition, 
the issue of production of the 2.223 MeV gamma-ray line from the capture
of neutrons onto protons and the possibility of its detection. 
This line has been observed in solar flares, the only astrophysical
site where neutron production has been observationally verified. Assuming
that each neutron results in a 2.223 MeV photon the flux computed in the
previous sections (Tables 2a and 2b) can be used to estimate the associated 
2.223 MeV  photon production rate by simply multiplying the associated 
production rate of neutrons by the solid angle subtended by the companion 
of the compact object. It is estimated that the latter is of order $\simeq 
10^{-3}$, leading to a flux of only $10^{36}-10^{37}$ photons per second
for an compact object of $\simeq 1$ solar mass; 
for a source at a distance of 1 kpc this corresponds to a flux
of $10^{-8}-10^{-7}$ cm$^{-2}$ sec$^{-1}$, too low to be observable with
near future instrumentation (INTEGRAL). However, for a compact 
object of larger size, the flux of the 2.223 MeV line could be observable;
for instance, assuming $m = 10, \; \alpha = 0.3$ and $\dot m =1 $,
equations (3) and (5) yield a luminosity  $L \sim 10^{38.5} \; {\rm erg 
\; sec} ^{-1}$  and a neutron production rate of $R \sim 10^{42.5} \; 
{\rm sec}^{-1}$. Assuming the same solid angle of interception by the 
companion as above ($\simeq 10^{-3}$), this figure yields the respectable 
flux of $\sim 3 \times 10^{-5}$ photons cm$^{-2}$ sec$^{-1}$, suggesting
that an observation of this line may not be totally improbable with 
the instruments aboard the INTEGRAL mission. 

A potential enhancement of the 2.223 MeV flux beyond the figures given 
above would be possible, if at larger distances the density profile 
of the accretion flow were to become less steep than the $r^{-3/2}$ 
dependence of ADAF 
and result to a column density sufficiently high to allow a fraction of the 
escaping neutrons to be deposited, cooled and captured within the flow in 
collisions with the accreting gas.  
Under such circumstances, most of the escaping neutrons could be 
used in the production of 2.223 MeV photons, increasing the probability
of detecting this nuclear line.   

\begin{acknowledgments}
We would like to thank Reuven Ramaty, Chris Shrader, Benzion Kozlovsky, 
and Allen Sweigart for a number of helpful discussions and comments 
pertaining to the present work, as well as the referee for several 
useful suggestions. NG  would also like to thank Reuven Ramaty for the 
hospitality at GSFC where most of this was performed. 
\end{acknowledgments}

\vfill\eject

\centerline{\bf TABLE 1}
\medskip
\centerline{\bf Nuclear Reactions in the Disk and their Threshold Energies}
\vskip 0.2 in
\hrule
\vskip 0.15 in
\halign{\qquad\hfil#\qquad\hfil&#\hfil&#\qquad\hfil&\hfil#\hfil\cr
	Reaction Number &Nuclear &Reactions& Threshold Energies [MeV]\cr
	\hrulefill  &  \hrulefill &\hrulefill       &   \hrulefill     \cr
1.1&p+ $\alpha$            &$\longrightarrow$ ${}^3$He + d     & 18.35\cr
1.2&\phantom{$\alpha$ + p} &$\longrightarrow$ ${}^3$H + p + p  & 19.81\cr
1.3&\phantom{$\alpha$ + p} &$\longrightarrow$ ${}^3$He + p + n & 20.58\cr
   &                       &                                  &      \cr
2.1a& $\alpha + \alpha$ &$\longrightarrow$ ${}^7$Be + n &18.99\cr
2.1b&\phantom{$\alpha + \alpha$} &$\longrightarrow$ ${}^7$Be$^*$ + n &
19.42\cr
2.2a&\phantom{$\alpha + \alpha$} &$\longrightarrow$ ${}^7$Li + p & 17.35\cr
2.2b&\phantom{$\alpha + \alpha$} &$\longrightarrow$ ${}^7$Li$^*$ + p
&17.83\cr
2.3a&\phantom{$\alpha + \alpha$} &$\longrightarrow$ ${}^6$Li + p + n
&24.60\cr
2.3b&\phantom{$\alpha + \alpha$}&$\longrightarrow$ ${}^6$Li$^*$ + p +
n&28.16\cr
2.4&\phantom{$\alpha + \alpha$} &$\longrightarrow$ $\alpha +{}^3$He +
n&20.58\cr
2.5&\phantom{$\alpha + \alpha$} &$\longrightarrow$ $\alpha +{}^3$H + p
&19.81\cr
2.6&\phantom{$\alpha + \alpha$} &$\longrightarrow$ $\alpha$ + d + d
&23.85\cr
   &                            &                                 &     \cr
3.1&$\alpha$ + n           &$\longrightarrow$ ${}^3$H + d     & 17.59\cr
3.2&\phantom{$\alpha$ + n} &$\longrightarrow$ ${}^3$H + p + n  & 19.81\cr
3.3&\phantom{$\alpha$ + n} &$\longrightarrow$ ${}^3$He + n + n  & 20.58\cr
   &                            &                                   &    \cr

4.1&p + ${}^7$Li           &$\longrightarrow$ $\alpha + \alpha$    & -17.35\cr
4.2&\phantom{p + ${}^7$Li} &$\longrightarrow$ ${}^7$Be + n      & 1.65\cr
   &                            &                                   &     \cr
5.1&n + ${}^7$Be            &$\longrightarrow$ $\alpha + \alpha$  & -18.99\cr
5.2&\phantom{n + ${}^7$Be} &$\longrightarrow$ ${}^7$Li + p       & -1.65\cr
   &                            &                                   &    \cr
6.1&p + ${}^6$Li          &$\longrightarrow$ ${}^4$He + ${}^3$He    & -4.02\cr
6.2&n + ${}^6$Li          &$\longrightarrow$ ${}^4$He + ${}^3$H     & -4.78\cr
   &                            &                                   &     \cr
7.1&p + ${}^3$He          &$\longrightarrow$ d + p + p    & 5.49\cr
7.2&\phantom{p+ ${}^3$He} &$\longrightarrow$ 3p + n       & 7.72\cr
   &                            &                                  &     \cr
8.1a&$\alpha + {}^3$He          &$\longrightarrow$ ${}^6$Li + p     & 4.02 \cr
8.1b&\phantom{$\alpha +{}^3$He} &$\longrightarrow$ ${}^6$Li$^*$ + p &
7.58\cr
8.2&\phantom{$\alpha +{}^3$He} &$\longrightarrow$ ${}^3$He+${}^3$He+n
&19.81\cr
8.3&\phantom{$\alpha +{}^3$He} &$\longrightarrow$ ${}^3$He+${}^3$H+p &
20.58\cr
8.4&\phantom{$\alpha +{}^3$He} &$\longrightarrow$ ${}^3$He+ d + d &
23.85\cr
   &                            &                                   &     \cr
9.1&p + ${}^3$H          &$\longrightarrow$ ${}^3$He + n   & 0.76 \cr
9.2&\phantom{p+ ${}^3$H} &$\longrightarrow$ d + d          & 4.03\cr
   &                       &                                   &    \cr
10.1a&$\alpha + {}^3$H          &$\longrightarrow$ ${}^6$Li + n   & 4.78\cr
10.1b&\phantom{$\alpha +{}^3$H} &$\longrightarrow$ ${}^6$Li$^*$ + n &
8.35\cr
10.2&\phantom{$\alpha +{}^3$H} &$\longrightarrow$ ${}^3$He+${}^3$H+n &
20.58\cr
10.3&\phantom{$\alpha +{}^3$H} &$\longrightarrow$ ${}^3$H+${}^3$H+p &
19.81\cr
10.4&\phantom{$\alpha +{}^3$H} &$\longrightarrow$ ${}^3$H+ d + d &
23.85\cr
   &                       &                                   &    \cr
11.1&$\alpha$ + d           &$\longrightarrow$ ${}^3$He + ${}^3$H  & 14.32\cr
11.2&\phantom{$\alpha$ + d} &$\longrightarrow$ $\alpha$ + p + n    & 2.22\cr
   &                            &                                 &     \cr
12.1&n + ${}^3$He          &$\longrightarrow$ ${}^3$H + p     & -0.76 \cr
   &                            &                                 &     \cr
13.1& p + n           &$\longrightarrow$ d + $\gamma$          & --2.22\cr}

\vskip 0.2 in
\hrule

\vfill\eject

\centerline{\bf TABLE 2a}
\medskip
\centerline{\bf FLUXES OF NEUTRONS (for an initial 90\% H and 10\% He
Composition)}
\centerline{\bf (neutrons \, s$^{-1}$)}
\vskip 0.2 in
\hrule
\vskip 0.15 in
\halign{\qquad\qquad\qquad\hfil#\qquad\hfil&#\hfil&\qquad\qquad#\hfil\cr
 $\dot M$ ($M_{\odot}$/yr)  &  $\alpha = 0.3$   &   $\alpha = 0.1$     \cr
			   &                   &                      \cr
	$10^{-12}$       &  $1.5\times 10^{35}$ &   $3.5\times 10^{36}$  \cr
	$10^{-11}$       &  $4.0\times 10^{37}$ &   $3.0\times 10^{38}$  \cr
	$10^{-10} $       &  $3.3\times 10^{39}$  &  $1.8\times 10^{39}$  \cr
 $3\times 10^{-10}$       &  $7.6\times 10^{39}$ &   $4.9\times 10^{39}$  \cr
	$10^{-9} $       &  $4.0\times 10^{39}$ &   $1.6\times 10^{40}$ \cr
 $3\times 10^{-9}$       &  $1.2\times 10^{39}$ &   $7.8\times 10^{39}$  \cr
        $10^{-8} $       &  $8.7\times 10^{38}$ &   $3.2\times 10^{39}$ \cr}

\vskip 0.2 in
\hrule

\vskip 2 truecm
\centerline{\bf TABLE 2b}
\medskip
\centerline{\bf FLUXES OF NEUTRONS (for an initial 10\% H and 90\% He
Composition)}
\centerline{\bf (neutrons \, s$^{-1}$)}
\vskip 0.2 in
\hrule
\vskip 0.15 in
\halign{\qquad\qquad\qquad\hfil#\qquad\hfil&#\hfil&\qquad\qquad#\hfil\cr
 $\dot M$ ($M_{\odot}$/yr)  &  $\alpha = 0.3$   &   $\alpha = 0.1$     \cr
			   &                   &                      \cr
        $10^{-12}$       &  $6.6\times 10^{36}$ &   $5.4\times 10^{37}$  \cr
        $10^{-11}$       &  $5.8\times 10^{38}$ &   $1.8\times 10^{39}$  \cr
        $10^{-10}$        &  $1.8\times 10^{40}$  &  $2.2\times 10^{40}$  \cr
 $3\times 10^{-10}$       &  $7.1\times 10^{40}$ &   $6.0\times 10^{40}$  \cr
        $10^{-9}$        &  $2.1\times 10^{41}$ &   $2.0\times 10^{41}$ \cr
 $3\times 10^{-9}$       &  $9.4\times 10^{40}$ &   $9.7\times 10^{40}$  \cr

        $10^{-8}$        &  $4.1\times 10^{40}$ &   $5.2\times 10^{40}$ \cr}

\vskip 0.2 in
\hrule

\vfill\eject
\centerline{\bf TABLE 3a}
\medskip
\centerline{\bf Final $^7$Li / Neutron Abundances in the Disk}
\centerline{\it (for an initial 90\% H and 10\% He Composition)}
\vskip 0.2 in
\hrule
\vskip 0.15 in
\halign{\qquad\qquad\hfil#\hfil\qquad&\hfil#\hfil&\qquad\qquad\hfil#\hfil\cr
 $\dot M$ ($M_{\odot}$/yr)  &  $\alpha = 0.3$   &   $\alpha = 0.1$     \cr
			   &                   &                      \cr
        $10^{-12}$       &  $8.6\times 10^{-6}$ / $4.1\times 10^{-4}$ &
		$6.0\times 10^{-5}$ / $1.0\times 10^{-2}$  \cr
        $10^{-11}$       &  $5.8\times 10^{-5}$ / $1.1\times 10^{-2}$ &
		$9.1\times 10^{-5}$ / $7.9\times 10^{-2}$  \cr
        $10^{-10}$        &  $8.8\times 10^{-5}$ / $7.7\times 10^{-2}$ &
		$1.4\times 10^{-4}$ / $2.3\times 10^{-1}$  \cr
 $3\times 10^{-10}$       &  $8.7\times 10^{-5}$ / $7.6\times 10^{-2}$ &
	 $1.3\times 10^{-4}$ / $2.1\times 10^{-1}$  \cr
        $10^{-9}$        &  $5.3\times 10^{-5}$ / $4.3\times 10^{-2}$ &
                 $7.2\times 10^{-5}$ / $1.4\times 10^{-1}$ \cr
 $3\times 10^{-9}$       &  $4.0\times 10^{-5}$ / $3.1\times 10^{-2}$ &
         $5.9\times 10^{-5}$ / $1.1\times 10^{-1}$  \cr
        $10^{-8}$        &  $2.9\times 10^{-5}$ / $2.9\times 10^{-2}$ &
                 $4.2\times 10^{-5}$ / $8.5\times 10^{-2}$ \cr}

\vskip 0.2 in
\hrule

\vskip 3 truecm
\centerline{\bf TABLE 3b}
\medskip
\centerline{\bf Final $^7$Li / Neutron Abundances in the Disk}
\centerline{\it (for an initial 10\% H and 90\% He Composition)}
\vskip 0.2 in
\hrule
\vskip 0.15 in
\halign{\qquad\qquad\hfil#\hfil\qquad&\hfil#\hfil&\qquad\qquad\hfil#\hfil\cr
 $\dot M$ ($M_{\odot}$/yr)  &  $\alpha = 0.3$   &   $\alpha = 0.1$     \cr
			   &                   &                      \cr
        $10^{-12}$       &  $7.2\times 10^{-4}$ / $1.9\times 10^{-2}$ &
		$4.6\times 10^{-3}$ / $1.2\times 10^{-1}$  \cr
        $10^{-11}$       &  $5.0\times 10^{-3}$ / $1.4\times 10^{-1}$ &
		$1.7\times 10^{-2}$ / $2.7\times 10^{-1}$  \cr
        $10^{-10}$        &  $1.9\times 10^{-2}$ / $2.8\times 10^{-1}$ &
		$3.2\times 10^{-2}$ / $3.9\times 10^{-1}$  \cr
 $3\times 10^{-10}$       &  $1.9\times 10^{-2}$ / $2.8\times 10^{-1}$ &
	 $3.1\times 10^{-2}$ / $3.7\times 10^{-1}$  \cr
        $10^{-9}$        &  $1.1\times 10^{-2}$ / $1.7\times 10^{-1}$ &
                 $1.7\times 10^{-2}$ / $1.9\times 10^{-1}$ \cr
 $3\times 10^{-9}$       &  $8.3\times 10^{-3}$ / $1.3\times 10^{-1}$ &
         $1.4\times 10^{-2}$ / $1.5\times 10^{-1}$  \cr
        $10^{-8}$        &  $6.1\times 10^{-3}$ / $9.8\times 10^{-2}$ &
                 $9.9\times 10^{-3}$ / $1.0\times 10^{-1}$ \cr}

\vskip 0.2 in
\hrule

\vfill\eject
\centerline{\bf TABLE 4}
\medskip
\centerline{\bf FLUXES OF 0.478 + 0.429 MeV LINES}
\centerline{\bf (photons \, cm$^{-2}$ \, s$^{-1}$)}
\vskip 0.2 in
\hrule
\vskip 0.15 in
\halign{\qquad\qquad\qquad\hfil#\qquad\hfil&#\hfil&\qquad\qquad#\hfil\cr
 $\dot M$ ($M_{\odot}$/yr)  &  $\alpha = 0.3$   &   $\alpha = 0.1$     \cr
			   &                   &                      \cr
        $10^{-12}$       &  $7.6\times 10^{-11}$ &   $5.3\times 10^{-10}$  \cr
        $10^{-11}$       &  $5.7\times 10^{-9}$ &   $1.8\times 10^{-8}$  \cr
        $10^{-10}$        &  $1.8\times 10^{-7}$  &  $2.5\times 10^{-7}$  \cr
 $3\times 10^{-10}$       &  $6.7\times 10^{-7}$ &   $7.7\times 10^{-7}$  \cr
        $10^{-9}$        &  $2.5\times 10^{-6}$ &   $2.6\times 10^{-6}$ \cr
 $3\times 10^{-9}$       &  $5.8\times 10^{-6}$ &   $6.2\times 10^{-6}$  \cr
        $10^{-8}$        &  $3.7\times 10^{-6}$ &   $4.9\times 10^{-6}$ \cr}

\vskip 0.2 in
\hrule

\vfill\eject
\centerline{\bf TABLE 5}
\medskip
\centerline{\bf Neutron-Induced Spallation Reactions of C, N, O, and Ne}
\vskip 0.2 in
\hrule
\vskip 0.15 in
\halign{\qquad\qquad\qquad#\hfil&#\qquad\hfil&\hfil#\hfil\cr
	Nuclear         &Reactions      & Threshold Energies [MeV] \cr
	\hrulefill      &\hrulefill     &   \hrulefill             \cr
n + ${}^{12}$C             &$\longrightarrow$ n + 3$\alpha$      & 7.28\cr
\phantom{n + ${}^{12}$C}   &$\longrightarrow$ $\alpha$ + $^9$Be  & 5.70\cr
\phantom{n + ${}^{12}$C}   &$\longrightarrow$ $^3$He + $^{10}$Be  &
19.47\cr
\phantom{n + ${}^{12}$C}   &$\longrightarrow$ $^3$H + $^{10}$B    & 18.93\cr
\phantom{n + ${}^{12}$C}   &$\longrightarrow$ $^7$Li +$\alpha$ +$^2$H &
22.40\cr
   &                       &                                   \cr
n + ${}^{14}$N             &$\longrightarrow$ $\alpha$+ $^{11}$B    & 0.16\cr
\phantom{n + ${}^{14}$N}   &$\longrightarrow$ 2$\alpha$ + $^7$Li  & 8.82\cr
   &                       &                                     \cr
n + ${}^{16}$O             &$\longrightarrow$ n + 4$\alpha$     & 14.44\cr
\phantom{n + ${}^{16}$O}   &$\longrightarrow$ $\alpha$ + $^{13}$C  & 2.22\cr
\phantom{n + ${}^{16}$O}   &$\longrightarrow$ 2$\alpha$ + $^9$Be  &
12.86\cr
\phantom{n + ${}^{16}$O}   &$\longrightarrow$ $^7$Li + $^{10}$B  & 23.63\cr
   &                       &                                      \cr
n + ${}^{20}$Ne             &$\longrightarrow$ $\alpha$+ $^{17}$O    & 0.59\cr
\phantom{n + ${}^{20}$Ne}   &$\longrightarrow$ 2$\alpha$ + $^{13}$C  &
6.98\cr
\phantom{n + ${}^{20}$Ne}   &$\longrightarrow$ 3$\alpha$ + $^9$Be  &
17.60\cr
\phantom{n + ${}^{20}$Ne}   &$\longrightarrow$ $^7$Li +$\alpha$ +$^{10}$B
&28.36\cr
\phantom{n + ${}^{20}$Ne}   &$\longrightarrow$ $^7$Li + $^{14}$N
&16.74\cr}

\vskip 0.2 in
\hrule

\end{document}